\documentclass{emulateapj}
\usepackage{txfonts}
\usepackage{color}

\slugcomment{Accepted by \textsc{the Astrophysical Journal}}
\shorttitle{He Ignition on Accreting Neutron Stars}
\shortauthors{Peng \& Ott}
\newcommand{\unitspace}{\ensuremath{\,}}
\newcommand{\usp}{\unitspace}
\newcommand{\numberspace}{\ensuremath{\;}}
\newcommand{\nsp}{\numberspace}
\newcommand{\unitstyle}[1]{\ensuremath{\mathrm{#1}}}
\newcommand{\power}[2]{\ensuremath{{#1}^{#2}}}

\newcommand{\cm}{\unitstyle{cm}}
\newcommand{\gram}{\unitstyle{g}}

\newcommand{\second}{\unitstyle{s}}

\newcommand{\grampersquarecm}{\gram\usp\power{\cm}{-2}} 

\newcommand{\beq}{\begin{equation}}
\newcommand{\eeq}{\end{equation}}
\newcommand{\beqa}{\begin{eqnarray}}
\newcommand{\eeqa}{\end{eqnarray}}

\newcommand{\iso}[2]{\ensuremath{\mathrm{^{#1}#2}}}

\newcommand{\mdot}{\ensuremath{\dot{m}}}
\newcommand{\medd}{\ensuremath{\mdot_{\mathrm{Edd}}}}

\newcommand{\CP}{\ensuremath{C_P}}
\newcommand{\enuc}{\ensuremath{\varepsilon_\mathrm{nucl}}}
\newcommand{\ecool}{\ensuremath{\varepsilon_\mathrm{cool}}}
\newcommand{\gpcc}{\ensuremath{\gram\nsp\cm^{-2}}}
\newcommand{\gpscps}{\ensuremath{\gram\nsp\cm^{-2}\nsp\second^{-1}}}

\begin{document}

\title{Helium Ignition on Accreting Neutron Stars with a New Triple-$\alpha$ reaction rate}

\author{Fang Peng and Christian D. Ott}
\affil{Theoretical Astrophysics, California Institute of Technology, 1200 E California Blvd., M/C 350-17, Pasadena, CA 91125}
\email{fpeng@caltech.edu; cott@tapir.caltech.edu}

\begin{abstract}
We investigate the effect of a new triple-$\alpha$ reaction rate from \citet{ogata09} on helium ignition conditions on accreting neutron stars and on the properties of the subsequent type I X-ray burst. We find that the new rate leads to significantly lower ignition column density for accreting neutron stars at low accretion rates. We compare the results of our ignition models for a pure helium accretor to observations of bursts in ultra-compact X-ray binary (UCXBs), which are believed to have nearly pure helium donors. For $\mdot > 0.001\,\medd$, the new triple-$\alpha$ reaction rate from \citet{ogata09} predicts a maximum helium ignition column of $\sim 3 \times 10^{9}\,\gpcc$, corresponding to a burst energy of $\sim 4\times 10^{40}$ ergs. For $\mdot \sim 0.01\,\medd$ at which intermediate long bursts occur, the predicted burst energies are at least a factor of 10 too low to explain the observed energies of such bursts in UCXBs. This finding adds to the doubts cast on the triple-$\alpha$ reaction rate of \citet{ogata09} by the low-mass stellar evolution results of \citet{dotter09}.
\end{abstract}


\section{Introduction}\label{introduction}

The triple-alpha ($3\alpha$) reaction, $3\,\iso{4}{He} \rightarrow \iso{12}{C} + \gamma$,  is of fundamental importance for stellar evolution and explosive phenomena involving neutron stars and white dwarfs. The $3\alpha$ reaction rate is generally well understood at temperatures higher than $\sim 10^{8}$ K where the reaction is resonant. At low temperatures, the rate is dominated by non-resonant reactions and uncertainties exist.  Recently, \citet{ogata09} (OKK09) studied the $3\alpha$ reaction rate at low temperatures ($10^7\,{\rm K} < T < 10^9\, {\rm K}$) by directly solving the three-body Schr\"{o}dinger equation. They found tremendous enhancement of the $3\alpha$ rate compared to previous results, including the most recent rate of \citet{fynbo05} (Fy05)\footnote{The Fy05 rate, currently adopted by the JINA REACLIB database (http://groups.nscl.msu.edu/jina/reaclib/db/), is within a factor of 10 of the previous standard rates from \citet{caughlan88} and the standard rates in the NACRE compilation \citep{angulo99}.}.  Compared to Fy05, the OKK09 result is a factor of $10^{26}$ greater at $10^7$ K, $10^{6}$ times greater at $10^8$ K, and becomes equal at $T \ge 2.5 \times 10^{8}$K. 

The increased $3\alpha$ rate allows helium ignition to occur in a lower temperature and density environment. This has the potential to strongly affect a broad range of astrophysical phenomena, including, but not necessarily limited to, helium burning in stars and on accreting white dwarfs and neutron stars.  \citet{dotter09} have shown that the new rate affects significantly the evolutionary track of low-mass stars of 1 and $1.5\,M_\odot$, resulting in a shortened or even almost disappearing red giant phase; a result that is incompatible with observations. \citet{saruwatari10} have studied the effect of the new rate on the ignition of helium accreting C-O white dwarfs and found that off-center helium detonation rather than central carbon ignition may dominate the mechanism of type Ia supernovae, contrary to previous results. 

In this work, we investigate the effect of the new $3\alpha$ reaction rate on the helium ignition conditions on accreting neutron stars and on the properties of the subsequent type I X-ray burst (XRB). We find that the He ignition layer is rather sensitive to the choice of the $3\alpha$ reaction rate. The new rate leads to significantly lower ignition column density at low mass accretion rates. We present the results of our ignition models for pure helium accretors and compare them directly to observations of intermediate long bursts from ultra-compact X-ray binaries (UCXBs), which are believed to have nearly pure helium donors. We find that the new $3\alpha$ reaction rate predicts burst energies that are too low to accommodate the observed energy of $\sim 10^{41}$ ergs of intermediate long bursts. 

\section{Helium Ignition Conditions}\label{ignition} 

The energy generation rate of the $3\alpha$ reaction is given by
\beqa
\varepsilon_{3\alpha} & = & N_A r_{3\alpha} Q_{3\alpha} \ , \\
& = & \frac{1}{6} N_A \left(\frac{Y}{4}\right)^3 \rho^2 Q_{3\alpha} f_{\rm sc} \langle \alpha \alpha \alpha \rangle \ , 
\eeqa
where $Y$ is the mass fraction of He and $Q_{3\alpha} \simeq 7.275$ MeV is the energy release from the reaction. $r_{3 \alpha}$ is the $3\alpha$ reaction rate per unit volume and $\langle \alpha \alpha \alpha \rangle$ is the temperature-dependent part of the rate (in units of $\rm{cm^6\,s^{-1}\,g^{-2}}$). $f_{\rm sc}$ is the electron screening enhancement factor which is most important in the high density, low temperature regime. We calculate $f_{\rm sc}$ from the screening potentials by using $f_{\rm sc} = \exp(h_{\alpha\alpha} + h_{\alpha \rm Be})$. Here,  $h_{\alpha\alpha}$ and $h_{\alpha \rm Be}$ correspond to the screening potentials for each step of the $3\alpha$ reaction, $\iso{4}{He} + \iso{4}{He} \leftrightarrow \iso{8}{Be}$ and $\iso{4}{He} + \iso{8}{Be} \rightarrow \iso{12}{C} + \gamma$, respectively. For two species $i$ and $j$ that create a compound nucleus ($\{Z_{\rm c}, A_{\rm c}\} = \{(Z_i + Z_j), (A_i + A_j)\})$ in a reaction, $h_{ij} = f(\Gamma_i) + f(\Gamma_j) - f(\Gamma_{c})$.\footnote{Here, we apply the linear mixing rule (LMR) for a multi-component plasma. Recent studies have shown that the LMR is not accurate at the Debye limit at $\Gamma \rightarrow 1$ \citep{potekhin09a} and at $\Gamma >> 1$ \citep{potekhin09b}. The correction to the LMR is small in our interested range of $\Gamma$ and is ignored here.}  Here, $\Gamma_i = (Z_i e)^2/(a_i k_{\rm B} T)$ is the Coulomb coupling parameter for species $i$ with $a_i$ being the ion sphere radius. $f(\Gamma_i)$ is the Coulomb free energy per ion for a one-component plasma of species $i$ for which we adopt the analytical fit of \citet{potekhin.chabirer:eos_solid} that matches with Monte Carlo simulations.To examine how the He ignition curve depends on the choice of screening enhancement factor, we also include in our calculations a different formalism of screening enhancement taken from \citet{itoh90}.

When matter is accumulated on the surface of a neutron star, it is heated by gravitational energy release, nuclear reactions, and energy flux emanating from the deep neutron star crust. The temperature eventually becomes sufficiently high to lead to explosive burning. We determine the He ignition condition that is fulfilled when heating by nuclear burning becomes faster than radiative cooling, i.e., 
\beq
\left. \frac{\partial \varepsilon_{3\alpha}}{\partial T}\right |_P \ge \left. \frac{\partial \ecool}{\partial T}\right |_P,
\label{ignition.e}
\eeq
where $\ecool = \rho KT/y^2$ is a one-zone approximation of the local cooling rate. The temperature derivatives are taken at constant pressure $P=gy$. Here $g$ is the gravitational acceleration on the neutron star surface and $y$ is the column density, i.e., the accumulated mass per unit area.  $K$ is the thermal conductivity. We use the same equation of state and thermal conductivity as described in \citet{Peng2007SEDIMENTATION-A}.

Throughout this letter, we assume a canonical neutron star mass of $M=1.4\,M_{\odot}$ and a radius of $R = 10$ km. The surface gravitational redshift is $z = [1-2GM/(Rc^2)]^{-1/2} -1 = 0.31$ and the gravitational acceleration is $g=(GM/R^2)(1+z) = 2.43 \times10^{14}\,{\rm cm\,s^{-2}}$. 

\begin{figure}[t]
  \centering
  \includegraphics[width=3.3in,trim=0.0cm -0.5cm 0.0cm -0.5cm]{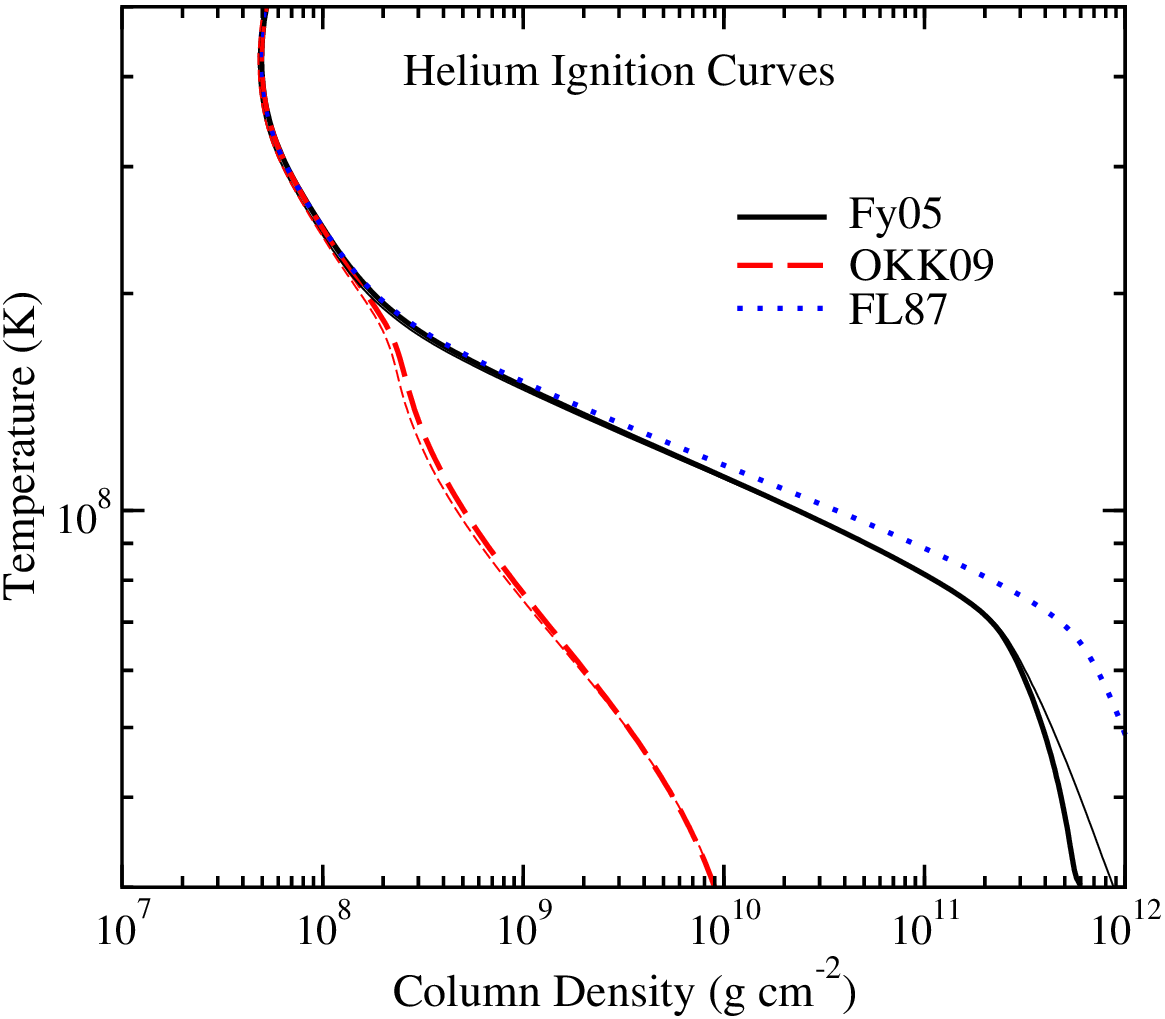}
  \caption{Ignition curve for a pure He neutron star accretor. Two $3\alpha$ reaction rates are adopted: Fy05 \citep{fynbo05} (\emph{solid} lines) and OKK09 \citep{ogata09} (\emph{dashed} lines). We multiply both rates by an electron screening enhancement factor, provided either by \citet{potekhin.chabirer:eos_solid} (\emph{thick} lines) or by \citet{itoh90} (\emph{thin} lines). The ignition curve with the analytical expression for the screened $3\alpha$ reaction rate from \citet{fushiki87} (\emph{dotted} line) is also shown for comparison.}
  \label{3a_ignition.f}
\end{figure}

To obtain the ignition column density for a given ignition temperature, we increase the column density until the ignition condition (eq.~[\ref{ignition.e}]) is reached. In Figure \ref{3a_ignition.f}, we compare the He ignition curves obtained with the $3\alpha$ reaction rates of Fy05 (\emph{solid} line) and OKK09 (\emph{dashed} line). We multiply both rates by an electron screening enhancement factor, provided either by \citet{potekhin.chabirer:eos_solid} (\emph{thick} lines) or by \citet{itoh90} (\emph{thin} lines). The results of the two screening formalisms agree quite well. Overall, the OKK09 rate leads to significantly lower ignition column density for the same ignition temperature. This effect is most prominent at temperatures below $\sim 2\times10^{8}$ K and, hence, for neutron stars accreting at low rates. Also shown in Fig.~\ref{3a_ignition.f} is the He ignition curve resulting from an analytical expression for the screened $3\alpha$ reaction rate given by \citet{fushiki87} (FL87, \emph{dotted} line). This rate has found broad application in the modeling of X-ray burst ignition conditions \citep[e.g.][]{cumming03,Cumming2005Long-Type-I-X-r,narayan.heyl:thermonuclear,cooper.narayan:theoretical}. The FL87 results agree quite well with those obtained with the Fy05 prescription. Significant differences appear only at low temperatures and high densities where both the uncorrected $3\alpha$ rates and electron screening corrections are different. 

\section{Ignition Models for Accreting Neutron Stars}\label{model}

Burst ignition and the properties of the subsequent burst depend on the accretion rate, the composition of the accreted matter, and on the interior properties of the accreting neutron star. The steady-state thermal structure of the accreted envelope is governed by the temperature and flux equations \citep[see][]{brown:nuclear},
\begin{eqnarray}
  \label{radiation.e}
  \frac{d T}{d y} &=& \frac{F}{\rho K}\ , \\
  \label{heat.e}
  \frac{d F}{d y} &=& 
  \CP\mdot\left( 
    \frac{dT}{dy} - \frac{T}{y}\nabla_\mathrm{ad}  \right)
   -  \enuc\ ,
\end{eqnarray}
where \CP\ is the specific heat at constant pressure and
$\nabla_\mathrm{ad} \equiv (\partial\ln T/\partial\ln P)_S$.  $\mdot$ is the local mass accretion rate, commonly normalized by the local Eddington mass accretion rate $\medd = 2 m_{\rm p} c/[(1+X) \sigma_{\rm TH} R]$ with $X$ being the mass fraction of hydrogen and $\sigma_{\rm TH}$ being the Thomson scattering cross section. For a pure He accretor, $\medd = 1.5 \times 10^5\,\gpscps$. $\enuc$ is the energy generation rate from nuclear burning. Here, we are considering only pure He accretion. The energy generation rate during the accumulation phase alone is therefore governed by the $3\alpha$ reaction, $\enuc = \varepsilon_{3\alpha}$. We adopt and compare the $3\alpha$ rates of Fy05, OKK09 and FL87. 

We integrate from the surface ($y=10^3\,\gpcc$) to the outer crust of the neutron star\footnote{The ignition point is not sensitive to the choice of the column density at the outer and  inner boundaries.}. At the surface, we use a radiative-zero condition, $F = 4/(3 \kappa y)\left[\sigma_{\rm R} T^4 - F_{\rm acc} \right]$, where $\kappa \approx 0.2\,(1+X)\,{\rm cm^{2}\,g^{-1}}$ is the opacity contributed by electron scattering, $\sigma_{\rm R}$ is the Stefan-Boltzmann constant, and $F_{\rm acc} \equiv GM\mdot/(2R)$ is the accretion flux. We employ a fuel/ash burning scheme and switch the composition from fuel to ash when the ignition condition (eq.~[\ref{ignition.e}]) is reached. The composition of the ash is set to be $\iso{56}{Fe}$, assuming complete burning\footnote{The composition of the ash has negligible effect on the thermal structure of the fuel layer.}. 

At the inner boundary, we set the flux $F = F_b \equiv Q_b\mdot$, where $Q_b$ is the energy transported outward from the deep crustal heating via electron capture, neutron emission, and pycnonuclear reactions \citep{sato79,haensel90a,haensel.zdunik:nuclear}. The total energy release from the crust is set to be $Q_{\rm crust} = 1.4$ MeV/nucleon, but not all of this will be transported to the surface. Some fraction of the heat will diffuse to the core, provided efficient core cooling by neutrinos. The exact value of the outward energy release $Q_b$ is sensitive to the core thermal structure which we do not include in our calculations. Several self-consistent ignition models that integrated deeper into the core and matched the core flux to the neutrino luminosity have been carried out \citep{brown:superburst, Cumming2005Long-Type-I-X-r, cooper.narayan:theoretical}. These models predict $Q_b \simeq 0.1$ MeV/nucleon at $\mdot \gtrsim 0.1\,\medd$ and increasing at lower accretion rates. However, we note that the precise value of $Q_b$ is sensitive to the poorly constrained neutrino emissivity in the core.  In this work, we take $Q_b$ as a free parameter, with the constraints provided by parameter studies of the neutrino cooling efficiency~\citep{Cumming2005Long-Type-I-X-r}. We iterate the integration process by adjusting the surface flux to match the inner boundary condition. This method is similar to that described in \citet{cumming03}. 

We neglect compressional heating $\sim \CP T \approx 0.02\,T_8$ MeV/nucleon, which is negligible compared to crustal heating. Also, for simplicity, we ignore the moderate energy release from He burning during accumulation ($\sim 0.03\,(\Delta Y/0.05)$ MeV/nucleon with $\Delta Y$ being the mass fraction of burned He prior to ignition)\footnote{If more than 10\% of the accumulated He is burned prior to ignition, the He burning energy release could play a role in heating the accumulated matter. This would lead to a smaller He ignition column density.}. Therefore, the thermal structure of the accumulated envelope is uniquely determined by the outward flux $F_b = 10^{21}\, {\rm erg \,cm^{-2}\, s^{-1}} \,(Q_b/0.1\nsp{\rm MeV}) (\mdot/10^4\,\gpscps)$, a product of $Q_b$ and $\mdot$.  For each $F_b$, we can determine the ignition column density $y_{\rm ign}$ by solving the thermal equations. Once $y_{\rm ign}$ is determined, we estimate the energy of the burst by $E_b = 4 \pi R^2 y_{\rm ign} Q_{\rm nuc}/(1+z) = 1.53 \times 10^{41}\, {\rm ergs} \, [y_{\rm ign}/(10^{10}\,\gpcc)]/(1+z)$. Here we assume that all of the accumulated fuel is burned in the burst and that the energy release from complete He burning up to the iron group is $Q_{\rm nuc} = 1.6$ MeV/nucleon. 

\begin{figure}[t]
  \centering
  \includegraphics[width=3.4in,trim=0.0cm 0cm 0cm -0.5cm]{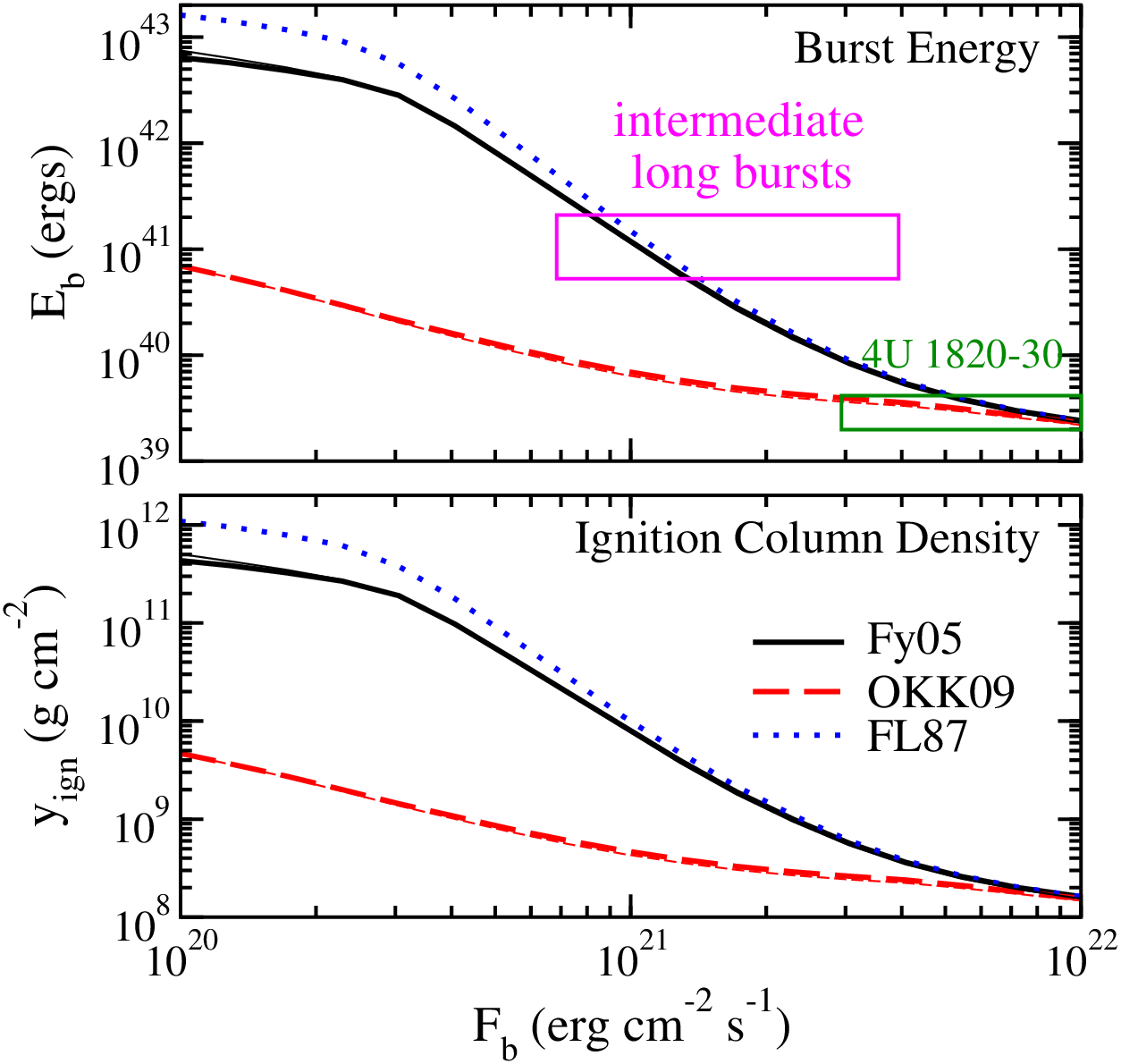}
  \caption{Burst energy and ignition column density as a function of the outward flux from the neutron star crust. The boxes mark the regions where the theoretically predicted burst energy matches the observed energy for 4U 1820-30 (\emph{green} color) and intermediate long bursts (\emph{magenta} color). The line styles are the same as in Fig. ~\ref{3a_ignition.f}.}
  \label{energy_yign_Fb.f}
\end{figure}

In Figure~\ref{energy_yign_Fb.f}, we show the ignition column density and the energy of the burst as a function of $F_b$. The new $3\alpha$ rate of OKK09 significantly reduces both the ignition column density and the burst energy, especially for accreting neutron stars with low crustal heat fluxes (at fixed $Q_b$, this corresponds to low mass accretion rates). The burst energy is reduced by a factor of 10 (100) at $F_b = 10^{21}\,(10^{20})\,{\rm erg \,cm^{-2}\, s^{-1}}$ to $\sim 7\times 10^{39}$ ergs ($7\times10^{40}$ ergs). 

\section{Application to Ultra-Compact X-ray Binaries} \label{sec:ucxbs}

Ultra-compact X-ray binaries (UCXBs), typically characterized by a low optical to X-ray flux ratio,  have very short orbital periods of less than $\sim 1$ hr. The small orbital separation implies that the companion is an evolved star that has lost its envelope and is quite hydrogen poor.  

4U 1820-30, located in the globular cluster NGC 6624, has an orbital period of $11.4$ minutes \citep{stella.ea:discovery} and may have a helium white dwarf secondary \citep{rappaport.ea:evolutionary}. It shows frequent type I X-ray bursts with total burst energy of $\approx 2-4\times10^{39}\, {\rm ergs}$, occurring at a local mass accretion rate $\mdot \approx 0.2\,\medd =  3 \times 10^{4}\,\gpscps$. The ratio of persistent fluence to burst fluence is $\alpha \approx 120$ \citep{haberl.ea:exosat}, suggesting burning of He-rich material. The crustal heating $Q_b$ at this mass accretion rate varies in a range of $0.1-0.4$ MeV/nucleon for a wide range of core neutrino emissivities based on the self-consistent theoretical ignition models of~\citet{Cumming2005Long-Type-I-X-r}. Therefore, $F_b \approx 3 - 10 \times 10^{21}\,{\rm erg\,cm^{-2}\,s^{-1}}$. The flux range together with the range of observed burst energies constrains the region where the theoretical predictions match observations. This region is marked by a green-colored box in Fig.~\ref{energy_yign_Fb.f}. All $3\alpha$ reaction rates can explain the observed energies, but the Fy05 and FL87 rates require high crustal heat fluxes. 

There are three UCXBs that have shown intermediate long bursts with durations of $10-40$ min and burst energies of $\approx 0.5-2\times10^{41}\,{\rm ergs}$, occurring at local mass accretion rates of $\mdot \approx 0.01\,\medd = 1.5\times10^3\,\gpscps$. Such intermediate long bursts have durations and energies intermediate between regular type I X-ray bursts and superbursts with durations longer than an hour. They may be due to explosive burning of a large pile of He slowly accumulated prior to the burst \citep{Cumming2005Long-Type-I-X-r,int-Zand2005On-the-possibil}.  Sources that have shown such bursts are SLX 1737-282~\citep{intzand02, falanga08}, SLX 1735-269~\citep{molkov05} and 2S 0918-549~\citep{int-Zand2005On-the-possibil}. SLX 1737-282 has to date shown exclusively intermediate long bursts with a recurrence time of $\sim 90$ days. It is also the only source that has shown more than one intermediate long burst.  

The uncertainty in the outward flux from crustal heating is dominated by the uncertainty of the mass accretion rate for intermediate long bursts. Given an uncertainty of a factor of 2 in $\dot{m}$, i.e., $\dot{m} \approx 750 - 3000\, \gpscps$, corresponding to a variation of $Q_b$ in the range of $0.9-1.4$ Mev/nucleon and $0.4-1.2$ MeV/nucleon at the minimum and maximum mass accretion rate~\citep[see][]{Cumming2005Long-Type-I-X-r}, the relevant range for the outward flux $F_b$ is $0.7-4 \times 10^{21}\,{\rm erg\,cm^{-2}\,s^{-1}}$. The magenta-colored box in Fig.~\ref{energy_yign_Fb.f} marks the constraint on the properties of intermediate long bursts. In the constrained range of outward flux, the new $3\alpha$ rate predicts burst energies $< 10^{40}$ ergs, which is much lower than the observed energies of intermediate long bursts. 

At very low mass accretion rates ($\mdot < 0.005\,\medd$), almost all of the deep crustal heating reaches the surface, i.e.,  $Q_b \simeq 1.4$ MeV/nucleon. This leads to $F_b = 2\times10^{20}\,{\rm erg\,cm^{-2}\,s^{-1}} (\mdot/0.001\,\medd)$. In this regime, the He ignition column density and predicted burst energy differ by orders of magnitude between the OKK09 and Fy05/FL87 $3\alpha$ rates. For instance, at $\mdot = 0.001\,\medd$, the Fy05 rate predicts a burst of $5 \times 10^{42}$ ergs (comparable to the energy of a superburst) while the OKK09 rate predicts a burst energy of only $\sim 4 \times 10^{40}$ ergs. He ignition at a deep column density of $y_{\rm ign} \sim 10^{11}\,\grampersquarecm$ could be a possible cause of superbursts \citep[see also][]{kuromizu03} and has been proposed as a possible explanation for a superburst from 4U 0614+091\citep{kuulkers10}. Therefore, any superburst detected at such low accretion rates can be used to constrain the $3\alpha$ reaction rate at low temperature and high densities. However, this kind of superburst is likely to be very rare since sources with such low accretion rates might not be able to sustain continuous accretion \citep[e.g.][]{lasota08}. Even if they are persistently accreting, it takes nearly $\sim 100$ years to accumulate enough fuel to trigger a superburst. 

\section{Discussions and Conclusions}\label{sec:disc-concl}

We have investigated the He ignition conditions on accreting neutron stars by considering a new $3\alpha$ reaction rate of \citet{ogata09} that is higher by orders of magnitudes at low temperatures than suggested by previous work. We demonstrate that the He ignition condition on neutron stars is sensitive to the choice of the $3\alpha$ reaction rate. The new rate leads to considerably smaller ignition column density and, hence, lower burst energy. For $\mdot > 0.001\,\medd$, the OKK09 rate predicts a maximum He ignition column density of $\sim 3 \times 10^{9}\,\gpcc$, corresponding to a burst energy of $\sim 4\times 10^{40}$ ergs. We have compared theoretical predictions of burst properties obtained with multiple $3\alpha$ rates to observations of intermediate long bursts in UCXBs, using the assumption that these systems are pure He accretors. We find that the new $3\alpha$ rate of \citet{ogata09} is inconsistent with observations and cannot explain the observed energy of such intermediate long bursts while calculations with the previous rates \citep[e.g.][]{fynbo05,fushiki87} yield predictions in general agreement with burst observations. 

Some binary evolutionary models predict that UCXBs might have a small amount of hydrogen ($X \sim 0.1$) in the accreted matter \citep[e.g.][]{podsiadlowski.ea:evolutionary}. Presence of hydrogen provides additional heating from hydrogen burning during accumulation which would lead to even smaller ignition column densities \citep{cumming03}. Hence, the $3\alpha$ rate of \citet{ogata09} would have an even harder time to explain intermediate long bursts in systems that are not pure He accretors. 

\acknowledgements

It is a pleasure to thank Edward Brown for helpful comments which improved the paper. We thank Evan O'Connor, Hendrik Schatz and James Truran for useful discussions and comments. We thank the anonymous referee for helpful suggestions. We also thank Kazuyuki Ogata for providing us with his triple-$\alpha$ reaction rates. This work is partially supported through NSF grants AST-0855535 and OCI-0905046.


\clearpage


\end{document}